\documentclass[final,3p,times]{elsarticle}

\usepackage{amssymb}
\usepackage{amsmath}
\usepackage{xcolor} 
\usepackage{lineno}
\usepackage{hyperref}
\usepackage{tabularx} 
\usepackage{multirow} 

\journal{Optics and Laser Technology} 

\begin{document}

\begin{frontmatter}



\title{Multimodal Fusion Network for Micro-displacement Measurement via Michelson Interferometer}


\author[1]{Zixing Jia\fnref{equal1}}
\author[1]{Jiawei Li\fnref{equal1}}
\author[1,2]{Xin Li\fnref{equal1}}
\author[1]{Zheng Kuang}
\author[1]{Yuehua Chen}
\author[1]{Ziping Chen\corref{cor1}}
\ead{chenzp6@mail.sysu.edu.cn}
\author[1,2]{Xudong Lin\corref{cor1}}
\ead{linxd39@mail.sysu.edu.cn}

\cortext[cor1]{Corresponding author}

\affiliation[1]{organization={School of Physics and Astronomy, Sun Yat-Sen University},
            city={Zhuhai},
            postcode={519082}, 
            country={China}}

\affiliation[2]{organization={MOE Key Laboratory of TianQin Mission, TianQin Research Center for Guangdong Gravitational Physics, Frontiers Science Center for TianQin, Gravitational Wave Research Center of CNSA, Sun Yat-sen University},
            city={Zhuhai},
            postcode={519082}, 
            state={Guangdong},
            country={China}}



\begin{abstract}
Single-wavelength Michelson interferometry offers high sensitivity but is limited by $\lambda/2$ phase ambiguity and strong noise sensitivity, hindering practical deployment. Conventional iterative fitting methods for phase recovery are computationally expensive and prone to convergence failure under severe noise, limiting throughput and reliability. We present a multimodal fusion network (MFN) for quasi-absolute micro-displacement measurement that is pretrained on simulated interferograms and fine-tuned with only $\sim$500 real images, enabling rapid adaptation to experimental conditions with minimal calibration effort. MFN fuses complementary interferometric observables and uses a dual-head design (regression for sub-$\lambda/2$ displacement and classification for integer-order) with an orthogonality regularizer to separate continuous and discrete components. Experimentally validated on a calibrated Michelson testbed, the method achieves nanometer-scale precision, strong robustness under severe mixed-noise conditions, an order-classification accuracy of 98\%, and real-time inference of $\sim$10 ms per frame. This hardware-efficient approach resolves single-wavelength ambiguity without additional optical channels or scanning, offering a practical, fast, and robust solution for interferometric metrology.
\end{abstract}


\end{frontmatter}


\section{Introduction}

Michelson interferometer, since its invention in the late 19th century, has become one of the most fundamental instruments in modern precision metrology owing to its high sensitivity, simple structure, and easily controllable optical path. It plays an indispensable role in various fields such as gravitational-wave detection \cite{abbott2016observation,bond2016interferometer}, precision metrology \cite{Wang:21} and materials science \cite{Chhaniwal:06}. However, the traditional Michelson interferometer architecture also exhibits inherent limitations. When the measured displacement exceeds half of the light wavelength (i.e., beyond $\lambda/2$), the interference signal suffers from a $2\pi$ phase ambiguity. Consequently, the displacement information obtained from the interference fringes or fitting equations becomes non-unique, leading to periodic errors or discontinuous phase jumps in the measurement results. To overcome this problem, several improved techniques have been developed, including heterodyne interferometry 
\cite{CHEN2023109821,chen2009absolute,TENG200716}, and wavelength scanning interferometry \cite{Psota:23,Ledl:17,Kredba:21}.

Neural networks can learn nonlinear mappings directly from data \cite{Hornik1989Multilayer}, capturing both spatial and temporal dependencies without relying on specific fitting formulas. This property is particularly beneficial for interferometric imaging, where signal degradation due to speckle, vibration, or defocus can severely hinder analytical reconstruction. Moreover, once trained, a neural network offers extremely fast inference \cite{Zhao:21}, allowing real-time or high-throughput displacement evaluation. 

Recent years have witnessed a rapid proliferation of deep learning methods in optical and photonic measurement fields. Notably, the extraction of phase information from interference fringes has seen significant advances in recent years. Deep learning–based fringe projection profilometry (FPP) \cite{Yu:20,Qian:20,Shen:24} methods now reconstruct the three-dimensional (3D) shape of target objects by extracting phase information from interference patterns. This method outperforms traditional phase-shifting techniques such as three-step phase-shifting (PS) and Fourier transform methods in both processing speed and noise robustness.

This trend motivates us to explore its application in Michelson interferometry for displacement measurement. In practical scenarios, interference fringes are inevitably contaminated by displacement-dependent systematic perturbation. For example, information related to object tilting. Compared to Least squares fitting approaches, our proposed deep learning–based scheme requires minimal additional hardware yet overcomes inherent displacement limitations. Furthermore, when applied to real-world images with strong noise, it achieves superior performance over standard fitting algorithms such as differential evolution (DE) \cite{storn1997differential} and quasi-Newton methods in both computational efficiency and noise resilience.

We develop a multimodal fusion network (MFN) that integrates spatial, spectral, and temporal cues \cite{yao2025deeplearningapproachspatiotemporal} from interferometric sequences to achieve robust and precise micro-displacement prediction. The proposed model jointly processes raw interferograms, their differential and frequency-domain representations, together with statistical temporal descriptors, through a unified data-driven framework. In contrast to previous fitting-based approaches, our method bypasses explicit phase unwrapping and directly regresses displacement magnitude with high precision and stability. Experimental results on data acquired from a modified Michelson interferometer demonstrate that the MFN remains robust and short fitting duration across diverse noise levels, representing a significant advancement toward practical, learning-based interferometric metrology.

\section{Methods}
\subsection{Overview of Michelson Interferometry Principle}

According to the basic principle of Michelson interference\cite{Hecht1998,huang2025diameter}, the intensity of the interference fringes is:
\begin{equation}
    I(x,y)= \tilde{I}_0(1+V\cos{\delta(x,y)})
\end{equation}
where \(\tilde{I_0}\) is the sum of the intensities of the two beams and \(V\)  is a constant with values ranging from 0 to 1. The information about the optical path difference is hidden in the phase \(\delta\). Considering the symmetry of the interference fringes, we can expand the dipole term of \(\delta\):
\begin{equation}
    \delta(x,y)=\phi +a r^2+ b r^4
\end{equation}
with the radius 
\begin{equation}
    r^2 = (x-x_0)^2+(y-y_0)^2
\end{equation}
where \(x_0\) and \(y_0\) are the positions of the centers of the interference fringes and a, b represent the coefficients. \(\phi \)   is the phase corresponding to the center of the stripe, including the interference order (\(\phi =2\pi m\)).

Since the laser used is a Gaussian beam (\(\tilde{I}_0=I_0e^{c [(x-x_c)^2+(y-y_c)^2]}\), where \(x_c\) and \(y_c\) are the coordinates of the point with the highest light intensity, and \(I_0\) and c are  are fitting parameters of this Gaussian beam.). Therefore, the final light intensity formula is (Model with background noise (\(k\)) added):
\begin{equation}\label{eq:simulate}
    I(x,y)= I_0 e^{c [(x-x_c)^2+(y-y_c)^2]}(1+V\cos(2\pi m +a r^2+b r^4)) +k
\end{equation}
To enhance robustness and convergence, the parameter estimation is performed via a two-stage optimization strategy, which we term the Heuristic Analytical Algorithm (HAA).

\begin{enumerate}
    \item \textbf{Global search.}  A differential evolution (DE) algorithm is used for robust global optimization over the parameter space. The initial population is generated using a Sobol \cite{sobol1967distribution} low-discrepancy sequence to ensure uniform coverage and avoid premature convergence.
    \item \textbf{Local refinement.} The best solution from DE is refined using gradient-based optimizers (L-BFGS-B \cite{zhu1997algorithm}) to achieve high-precision convergence.
\end{enumerate}
Prior to fitting, each image undergoes preprocessing:
\begin{itemize}
    \item Images are downsampled to 75\% of their original size using bilinear interpolation to accelerate computation.
    \item A small Gaussian blur ($\sigma = 0.5$ pixels) is applied to suppress high-frequency noise.
    \item Pixel intensities are normalized to the range $[0, 1]$.
\end{itemize}
 $m$ is used to compute displacement. According to the principle of Michelson interferometry, the change in displacement $\Delta L$ between consecutive frames relates to the phase change $\Delta m$ by
\begin{equation}
    \Delta L  =\Delta m \frac{\lambda}{2} = (n+\kappa)\frac{\lambda}{2}
\end{equation}
where $n$ represents the integer component of the fringe variation and $\kappa$ represents the fractional component. For conventional  iterative fitting methods(e.g.,HAA), the fractional component $\alpha$ can usually be determined with relatively high accuracy, whereas estimating the integer component $n$ remains more challenging. In the following sections, we introduce the proposed model(multimodal fusion network) and describe how it simultaneously estimates these two components.

\subsection{Model Architecture Overview}

\begin{figure}[ht]
    \centering
    \includegraphics[width=0.9\textwidth]{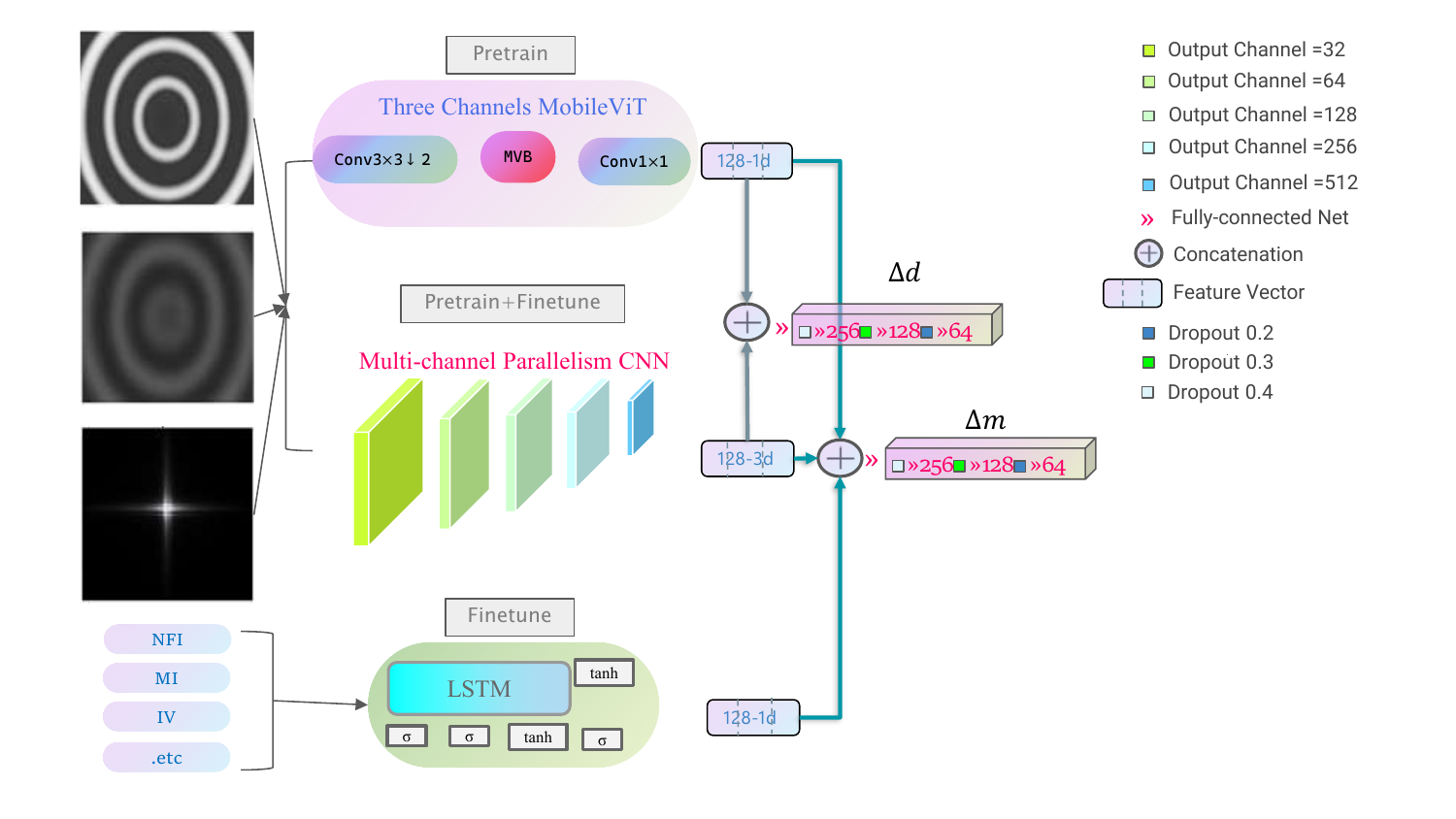}
    \caption{Architecture of the proposed multimodal fusion network.
    The model integrates: (1) three single-channel CNNs for raw, differential, and FFT-transformed interferograms (local features);
    (2) a MobileViT backbone for joint global feature extraction; and
    (3) an LSTM encoder for temporal numerical descriptors (activated only during fine-tuning).
    The multimodal feature vector branches into two parallel heads: 
    a displacement regression head that utilizes features from the two image-based branches (CNN + MobileViT) to achieve high-precision prediction of sub-$\lambda/2$ displacements, 
    and an order classification head that combines features from all three branches (CNN + MobileViT + LSTM) to learn the complex mapping from images and noise statistics to interference orders.
    Both heads are regularized by an orthogonality loss to ensure feature disentanglement.}
    \label{fig:architecture}
\end{figure}

We propose a multimodal fusion network that integrates both image-based and temporal statistical cues to predict nanoscale displacements with explicit order awareness. As illustrated in Fig~\ref{fig:architecture}, the overall framework retains a three-branch fusion design consisting of two image-based feature extraction pathways and one numerical modeling branch. On top of these, two parallel prediction heads are introduced: one for continuous displacement regression and another for discrete order classification. This design allows the model to disentangle fine-grained sub-period estimation from integer-order ambiguity inherent to interferometric measurements.

The first image branch comprises three parallel single-channel convolutional subnetworks, each processing a distinct visual modality: the raw interferogram (encoding spatial phase structures), the frame-difference image (capturing dynamic motion patterns), and the FFT-domain representation (reflecting global frequency distributions)~ \cite{simonyan2014twostreamconvolutionalnetworksaction,leethorp2022fnetmixingtokensfourier}. Each subnetwork produces a 128-dimensional feature vector that preserves modality-specific information without mutual interference. The second image branch employs a MobileViT~ \cite{mehta2022mobilevitlightweightgeneralpurposemobilefriendly} backbone that fuses the three modalities treated as a pseudo-RGB input. Through transformer-based self-attention, it models long-range spatial dependencies and cross-channel interactions, providing global contextual cues complementary to local CNN representations~ \cite{Gao2024Explainable}.

The third branch encodes a short temporal sequence of seven scalar descriptors (e.g., image entropy, fringe contrast, intensity variance) using a Long Short-Term Memory (LSTM) network~ \cite{LSTM}. This branch is activated only during the fine-tuning stage and is specifically designed to capture the statistical relationship between noise evolution and interference order transitions. In essence, it functions as a dedicated module that learns the complex mapping from noise patterns and temporal cues to discrete order categories within the interferometric system.

The outputs of all branches are concatenated into a 640-dimensional multimodal feature vector. From this joint representation, two prediction heads are constructed independently. The \emph{displacement head} receives features from the two image-based branches and performs regression over the continuous range $[0,\lambda/2]$, corresponding to the fractional optical path difference within one half-wavelength. The \emph{order head}, in contrast, receives the full concatenated representation (including the LSTM embedding) and classifies the interference order into one of five categories $\{-2,-1,0,1,2\}$. These two heads are trained jointly but maintain separate fully connected layers from the feature concatenation stage onward, avoiding feature entanglement \cite{Disentangle} that might otherwise arise from a shared latent representation. A soft orthogonality constraint \cite{cui2022multitaskaetorthogonaltangent,Phunruangsakao2022} between their final hidden representations further enforces statistical independence, preventing either branch from learning redundant or noise-correlated cues.

This modular design allows the network to simultaneously model continuous and discrete components of displacement, achieving robust performance across varying levels of interferometric noise and optical degradation.

\subsection{Feature Representation and Temporal--Spatial Encoding}

To enable displacement prediction from interferometric image sequences, we preprocess each frame into three image channels and one 7-dimensional numerical feature vector. This multi-modal design enriches the learning representation and injects both spatial and statistical cues into the regression model \cite{LeCun2015Deep}. The three image channels and the seven numerical descriptors are summarized in Table~\ref{tab:features}.

To accommodate the hybrid nature of our network where the numerical branch processes time-series sequences while the image branches operate on single-frame inputs, we implement a sliding-window strategy during data preparation. For each target frame \(I_t\), the image-branch inputs are produced from the single frame \(I_t\) (yielding $I_{\text{orig}}$, $I_{\text{diff}}$ and $I_{\text{fft}}$), while the numerical branch consumes a short temporal sequence of the scalar descriptors to capture temporal evolution.

Concretely, the numerical branch receives a window of 7-dimensional feature vectors
\begin{equation}
\{\mathbf{f}_{t-\tau},\dots,\mathbf{f}_t\},\qquad \mathbf{f}_s \in \mathbb{R}^7,
\end{equation}
where the window length is \(L_w=\tau+1\). In our experiments we choose \(L_w\in[2,5]\) (i.e., \(\tau\in[1,4]\)), which balances the need to capture short-term dynamics against computational cost and model stability. Shorter windows (e.g., near sequence starts) are zero-padded to match the batch maximum window length \(L_{\max}\) during collating. The collate function therefore emits:
\begin{equation}
\texttt{numerical\_seq\_batch} \in \mathbb{R}^{N \times L_{\max} \times 7},\qquad
\texttt{seq\_lengths} \in \mathbb{N}^{N},
\end{equation}

where \(N\) is the minibatch size. Here \(\texttt{seq\_lengths} = [\ell_1,\ldots,\ell_N]\) is an integer vector with each \(\ell_n\) indicating the true (unpadded) length of the \(n\)-th sequence in the batch; by construction \(1\le \ell_n \le L_{\max}\).

In the model forward pass the numerical batch and \texttt{seq\_lengths} are used together so the LSTM can correctly ignore padding. In practice we employ one of the following standard strategies:

\begin{itemize}
    \item \textbf{Packed sequences:}\\
    Use PyTorch's \texttt{pack\_padded\_sequence} before feeding the LSTM and \texttt{pad\_packed\_sequence} after, which requires sequences to be sorted by decreasing length.
    \item \textbf{Masking:} \\
    Process the padded tensor directly and apply an explicit mask derived from \texttt{seq\_lengths} when computing LSTM outputs / losses, thus zeroing contributions from padded time steps.
\end{itemize}

Both approaches prevent gradients and statistics from being contaminated by padding values and allow the LSTM to focus on real temporal information. After encoding, the numerical branch yields a fixed-dimensional embedding (e.g., by taking the last valid hidden state or via an attention/pooling over time), which is then concatenated with the image-branch embeddings to form the joint representation for regression.

This sliding-window design therefore captures short-term temporal dynamics in the numerical domain while keeping the image-branch computation frame-level and efficient. The combined representation benefits from both spatial detail (via per-frame CNN/MobileViT features) and temporal context (via the LSTM over numerical statistics), improving robustness when interferogram quality varies across frames.

\subsection{Training and optimization}

The network is trained in two stages. In the first stage, a large-scale set of simulated interferograms is used to pretrain the displacement regression branch. Although noise components are introduced into the simulated interferograms, the corresponding displacement values remain precisely known, enabling accurate supervision for regression learning. The simulated data include randomized phase terms, varying fringe contrast, and a range of additive noise components, enabling the model to learn the nonlinear mapping between interferogram structure and the fractional displacement within $[0,\lambda/2]$. During this pretraining phase, only the CNN and MobileViT branches are activated; the LSTM branch and the order-classification head remain uninitialized. The objective function is the mean squared error:
\begin{equation}
    L_{\text{pre}} = \frac{1}{N}\sum_{i=1}^{N} (\hat{d}_i - d_i)^2 .
\end{equation}

In the second stage, the model is fine-tuned using real interferometric sequences collected from the experimental setup in order to adapt the pretrained features to practical measurement conditions. Compared with simulated data, real interferograms contain complex noise sources. During this phase, the MobileViT backbone and the first several convolutional blocks in each CNN branch are frozen to preserve the pretrained spatial priors, while the LSTM branch and both prediction heads are optimized jointly. The displacement regression head and the order-classification head use separate loss functions to prevent interference between continuous and discrete prediction tasks. In particular, simulated data are mainly used to supervise the regression branch, as precise displacement labels are available, whereas real experimental data are primarily used to train the classification branch. This strategy reduces the model’s reliance on the calibration process and improves its transferability to different experimental conditions.

For the displacement branch, the optimization objective is
\begin{equation}
    L_{\text{Displacement}} = \alpha L_{\text{disp}} + \gamma L_{\text{orth}},
\end{equation}
where
\begin{equation}
    L_{\text{disp}} = \frac{1}{N}\sum_{i=1}^{N}(\hat{d}_i - d_i)^2
\end{equation}
and
\begin{equation}
    L_{\text{orth}}
    = \| h_s^\top h_n \|_F^2
\end{equation}
is an orthogonality regularizer that penalizes correlations between the batch-wise hidden representations $h_s$ (displacement head) and $h_n$ (order head), promoting feature decoupling. $|| \cdot ||_F$ denotes the Frobenius norm, ensuring that correlations across all feature dimensions are penalized.

Similarly, the order-classification head is optimized using
\begin{equation}
    L_{\text{Order}} = \beta L_{\text{order}} + \gamma L_{\text{orth}},
\end{equation}
where
\begin{equation}
    L_{\text{order}}
    = -\frac{1}{N}\sum_{i=1}^{N}\sum_{k} y_{i,k}\log(\hat{p}_{i,k})
\end{equation}
is the cross-entropy loss over the five interference-order categories $\{-2,-1,0,1,2\}$.

All experiments use the Adam optimizer~ \cite{kingma2017adammethodstochasticoptimization} with a learning rate of $1\times10^{-4}$ and weight decay of $5\times10^{-5}$. The coefficients are empirically set to $(\alpha,\beta,\gamma) = (1.0,\;0.5,\;0.05)$ to balance regression accuracy, classification performance, and feature orthogonality. This two-stage optimization scheme allows the model to retain the pretrained displacement prior while effectively adapting to experiment-specific noise and contrast characteristics.

\section{Experiments}

\subsection{Optical Interferometer Experiment}

Because our task is a supervised learning problem, it is necessary to obtain paired data consisting of samples and corresponding ground-truth labels. The purpose of this experiment is to generate large-scale interferometric fringe images and their associated displacement data for model training and validation.

\begin{figure}[htbp]
    \centering
    
    \begin{minipage}[t]{0.48\textwidth}
        \centering
        \includegraphics[width=\textwidth]{ 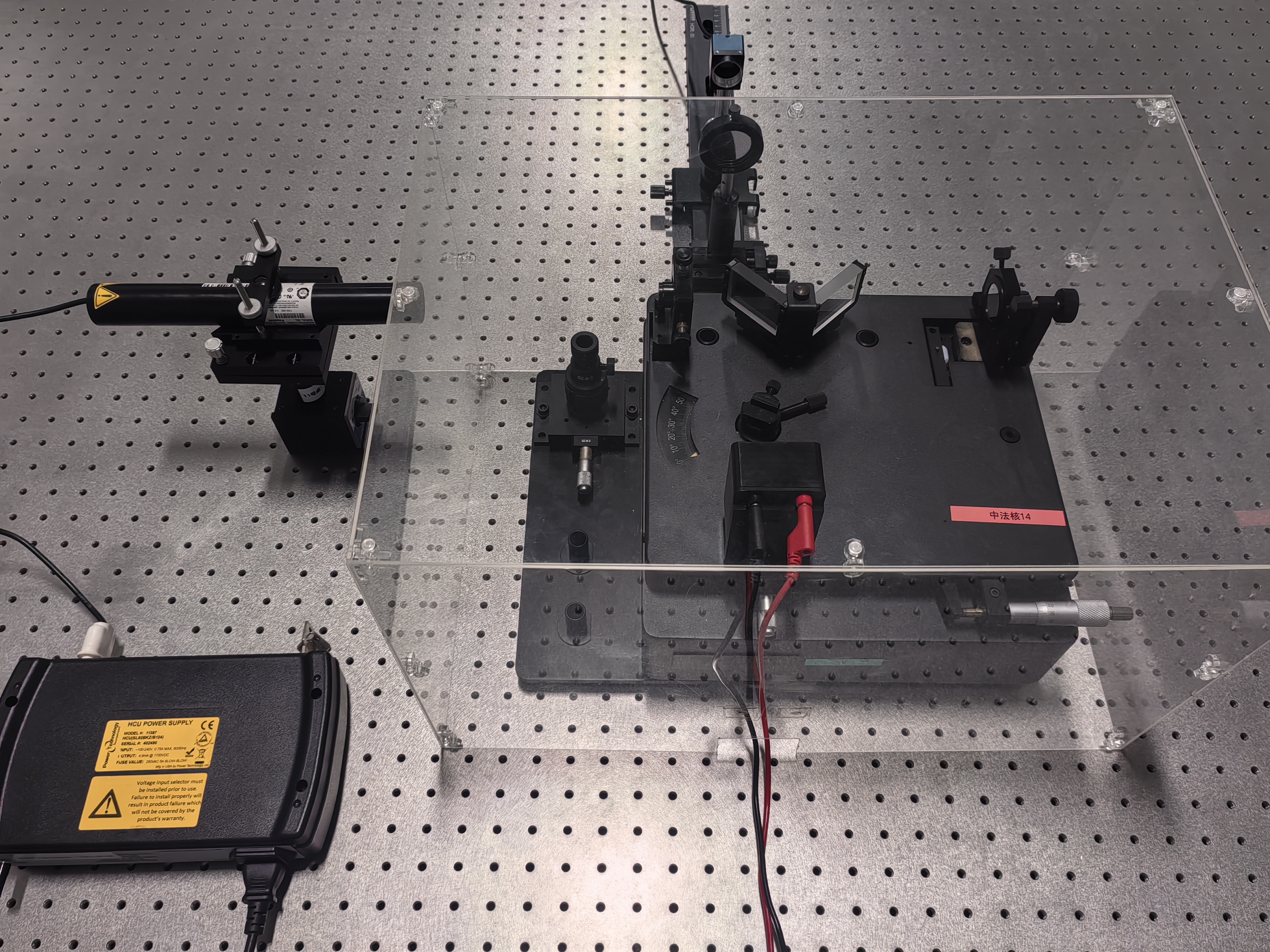}

    \end{minipage}
    \hfill
    \begin{minipage}[t]{0.48\textwidth}
        \centering
        \includegraphics[width=\textwidth]{ 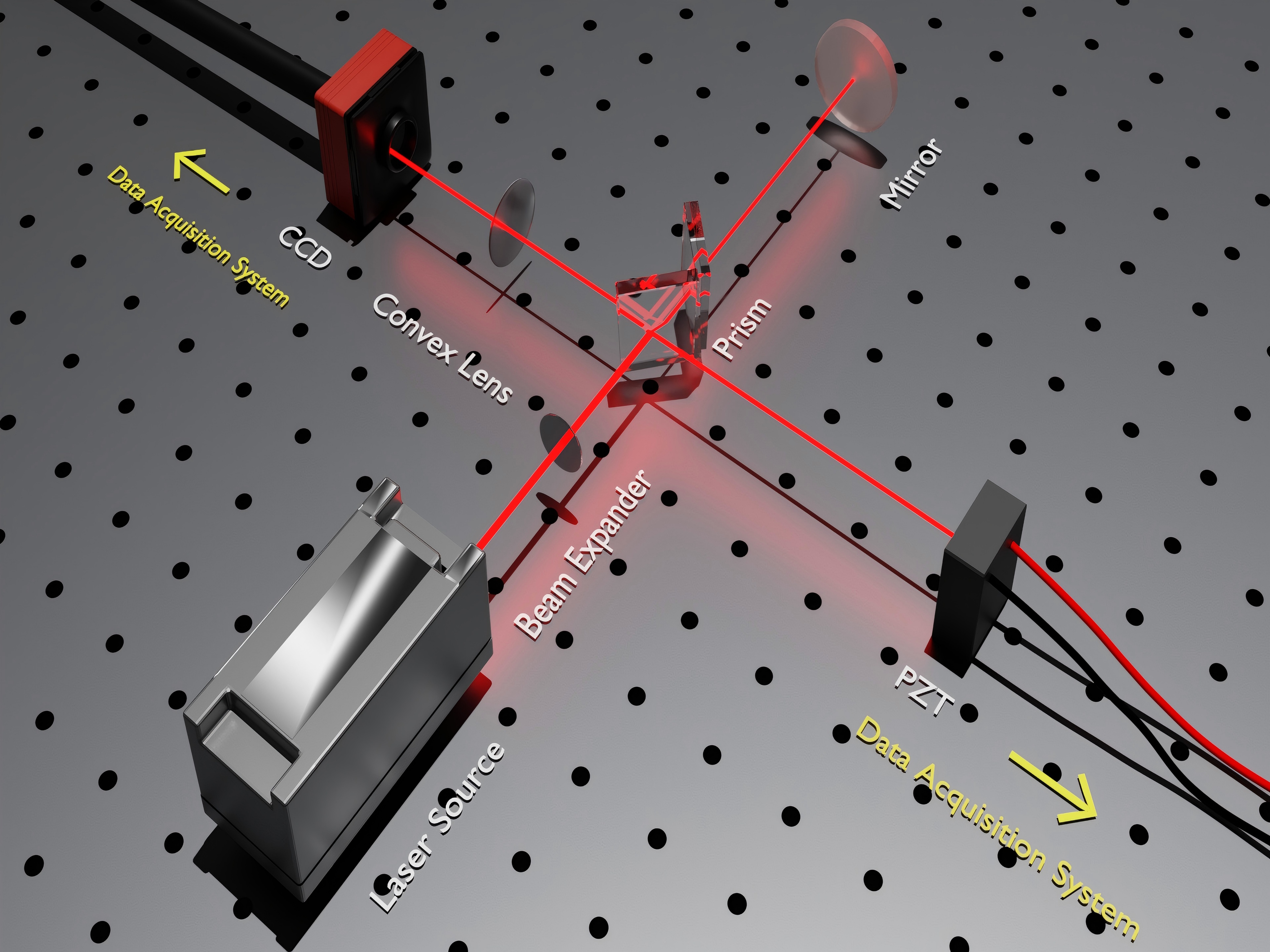}
\end{minipage}
    \caption{\textbf{(Left)} Photograph of the modified Michelson interferometer setup.\textbf{(Right)}  Schematic of the modified Michelson interferometer. Key components include: a He–Ne laser, beam expander and collimator, convex lens, prism-based beam splitter, movable mirror mounted on a piezoelectric actuator, fixed mirror, and CCD camera. Red lines indicate the optical path}
    \label{fig:michelson_combined}

\end{figure}

As illustrated in Fig.~\ref{fig:michelson_combined}, we developed a fully automated data acquisition system based on a modified Michelson interferometer \cite{merzouk_highly_2016}. The system mainly includes an interferometer (\textit{SGM-4 / Tianjin Gangdong}), a camera (\textit{Daheng MER-130-30UM}), and a voltage-controlled piezoelectric actuator (\textit{PST150/2×3/10 / XinTian Tech.}). By varying the driving voltage of the actuator, the relative displacement between the two interferometer arms changes, producing different interference patterns. During this process, both voltage and image data are recorded automatically and synchronously, forming a dataset that directly links interferograms to displacement values.

To eliminate human-introduced errors during measurement \cite{Babu2023Deep}, the acquisition is fully automated  through a custom control program, ensuring stable and repeatable measurements. To reduce environmental interference such as airflow and vibration, the optical path is enclosed with acrylic panels, effectively improving the stability and consistency of the captured interferograms.

Detailed specifications of all optical and electronic components are provided in the Supplemental Document Section 8.

\subsection{Voltage–Displacement Characterization of the Piezoelectric Actuator}

\begin{figure}[htbp]
    \centering
    
    \begin{minipage}[t]{0.48\textwidth}
        \centering
        \includegraphics[width=\textwidth]{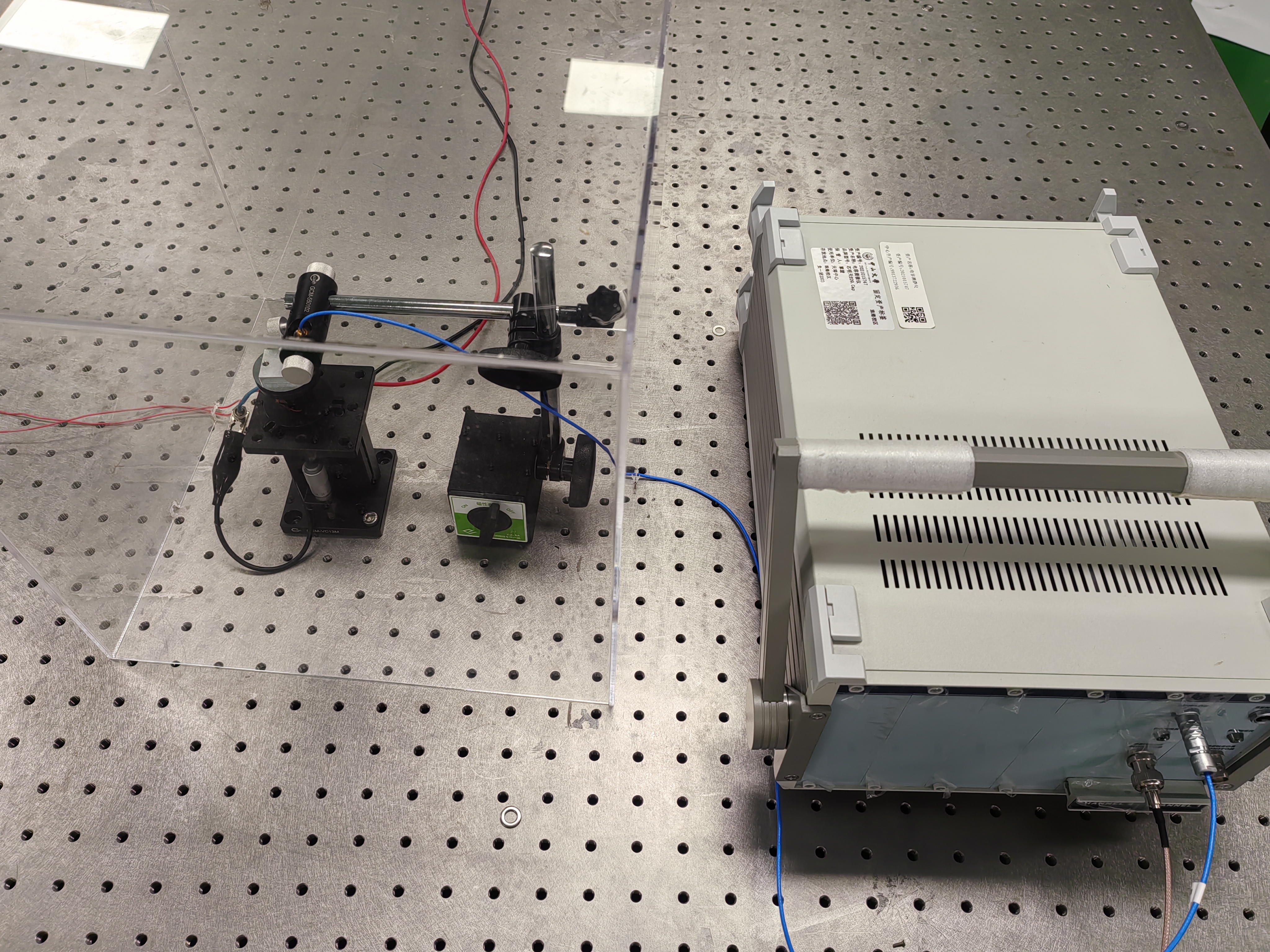}

    \end{minipage}
    \hfill
    \begin{minipage}[t]{0.48\textwidth}
        \centering
        \includegraphics[width=\textwidth]{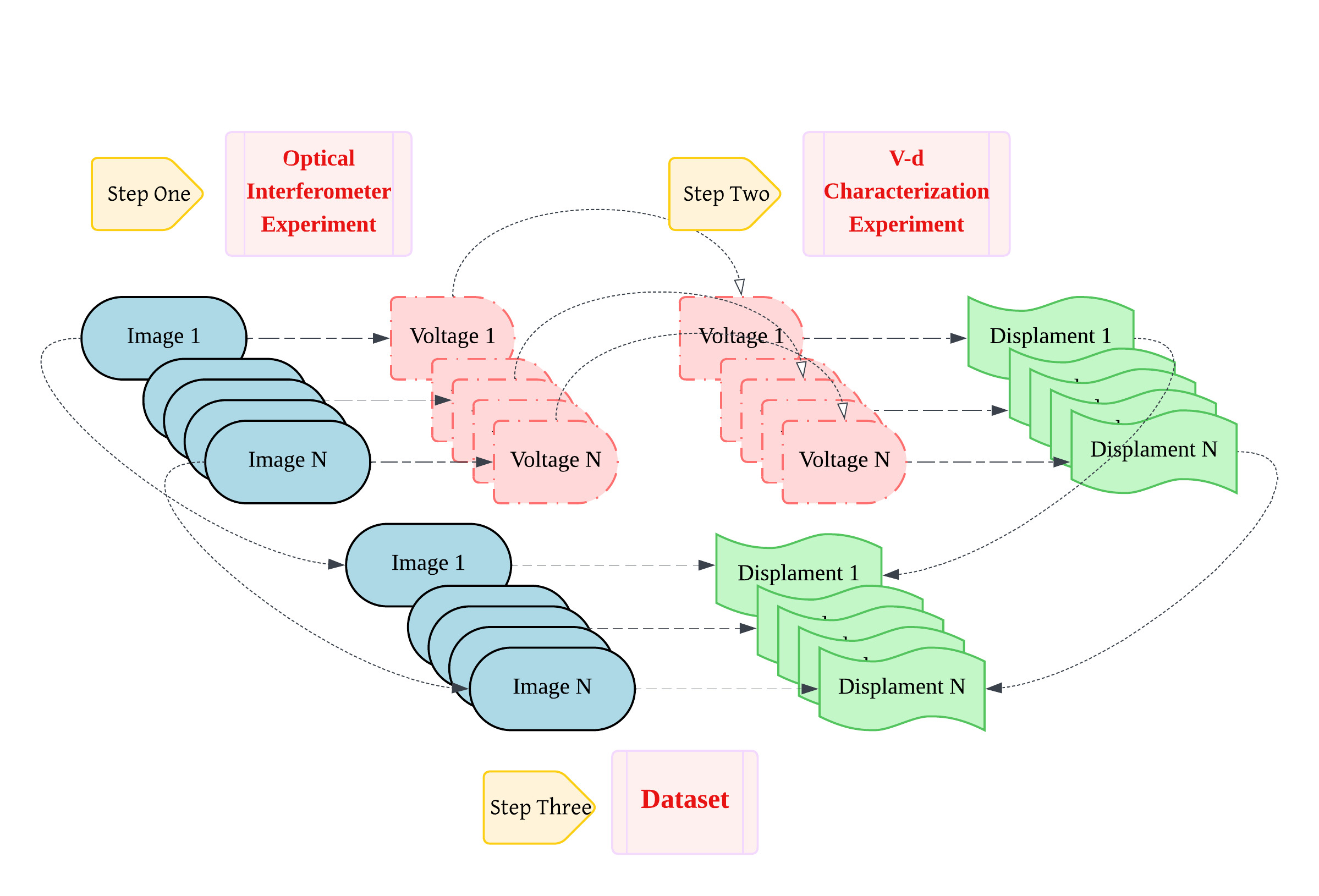}
    \end{minipage}
    \caption{\textbf{(Left)} Photograph of the voltage–displacement ($V$–$d$) characterization setup, showing the piezoelectric actuator mounted on the capacitive micrometer inside an acrylic enclosure. \textbf{(Right)} Two-stage calibration: synchronized interferogram acquisition with voltage inputs followed by displacement calibration, yielding paired image-displacement datasets for model training. Alternatively: Sequential interferogram-displacement calibration produces the aligned image-position datasets required for model development.}
    \label{fig:vd_setup}

\end{figure}

To obtain the labels required for supervised training, it is necessary to calibrate the voltage–displacement (\(V\)–\(d\)) relationship of the piezoelectric actuator. As shown in Fig.~\ref{fig:vd_setup}, this measurement is performed using a capacitive micrometer.

Considering that the piezoelectric actuator exhibits hysteresis and other nonlinear effects, it is essential to accurately measure its actual displacement response to the applied voltage. To achieve this, we developed an automated measurement system similar to the interferometric acquisition setup described in the previous section. This system controls the applied voltage to the actuator and synchronously records both voltage and displacement data from the capacitive micrometer.

To ensure consistency between the interferometric dataset and the displacement calibration, the amplitude of voltage variation and the dwell time at each voltage level are kept identical to those used in the interferometer experiment. This guarantees that the previously obtained image–voltage pairs can be reliably converted into image–displacement pairs through the measured \(V\)–\(d\) relationship.

Despite this calibration, several sources of uncertainty remain in the experimental measurements. First, the capacitive displacement sensor exhibits a static noise with a standard deviation of approximately 14 nm under stationary conditions, which propagates to about 20 nm when considering differential displacement between consecutive measurements. Second, during dynamic operation the measurement uncertainty increases due to sensor bandwidth limitations and actuator dynamics. In addition, environmental factors such as mechanical vibration, thermal drift, and minor discrepancies between the interferometer setup and the calibration configuration may further affect the stability of the displacement reference.Considering these factors, the overall background uncertainty of the experimental system can significantly influence the regression branch responsible for predicting sub-wavelength displacements. However, since the real experimental data are primarily used for training the classification branch, which operates on a much larger displacement scale ($\approx 315 nm$), its performance is relatively insensitive to this level of noise. A more detailed analysis is provided in the Supplemental Document Section 7.

\begin{figure}[ht]
    \centering
    \begin{minipage}[b]{0.45\textwidth}
        \centering
        \includegraphics[width=\textwidth]{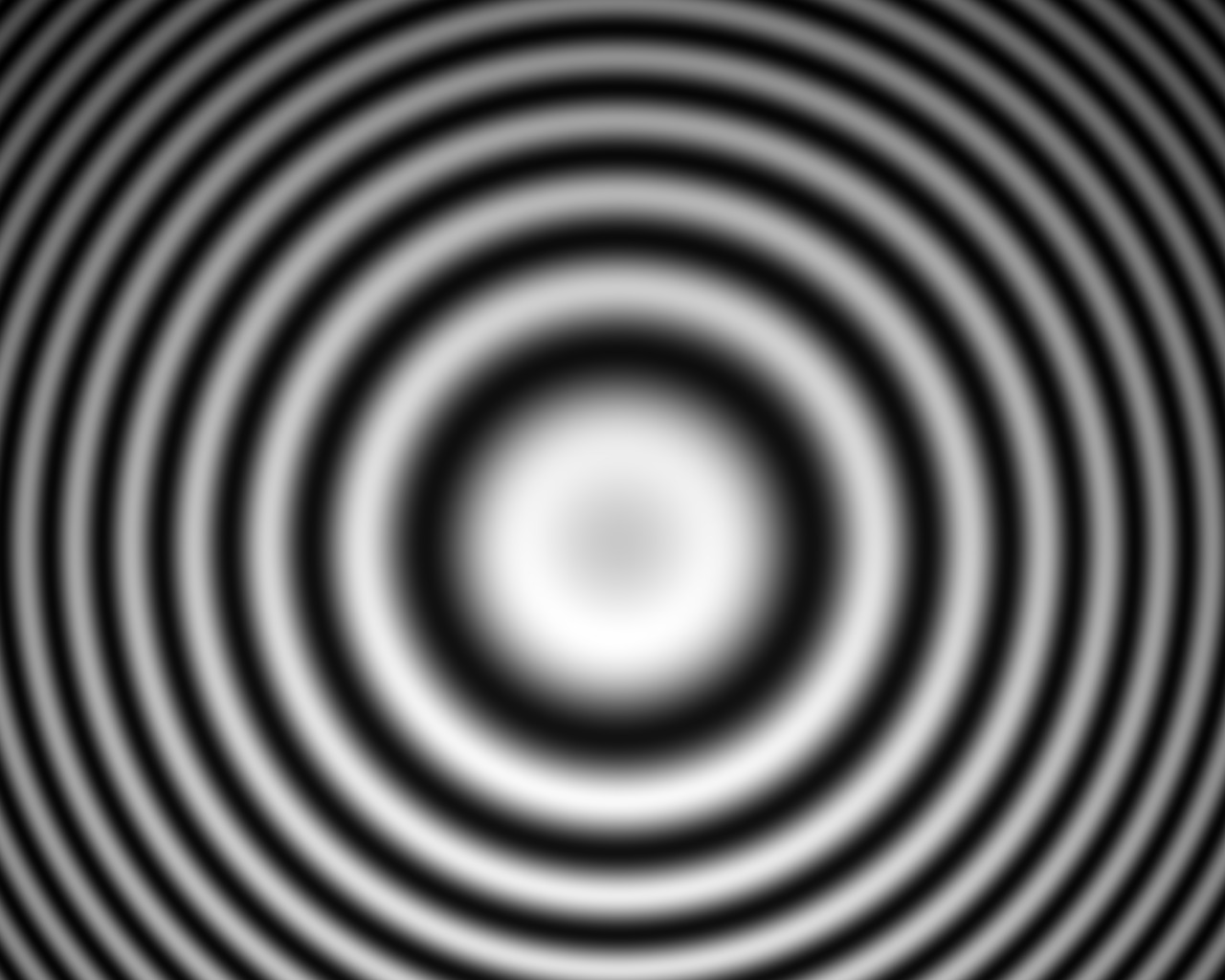}\\
        \small{\(\eta\) = 0.0}
    \end{minipage}
    \hfill
    \begin{minipage}[b]{0.45\textwidth}
        \centering
        \includegraphics[width=\textwidth]{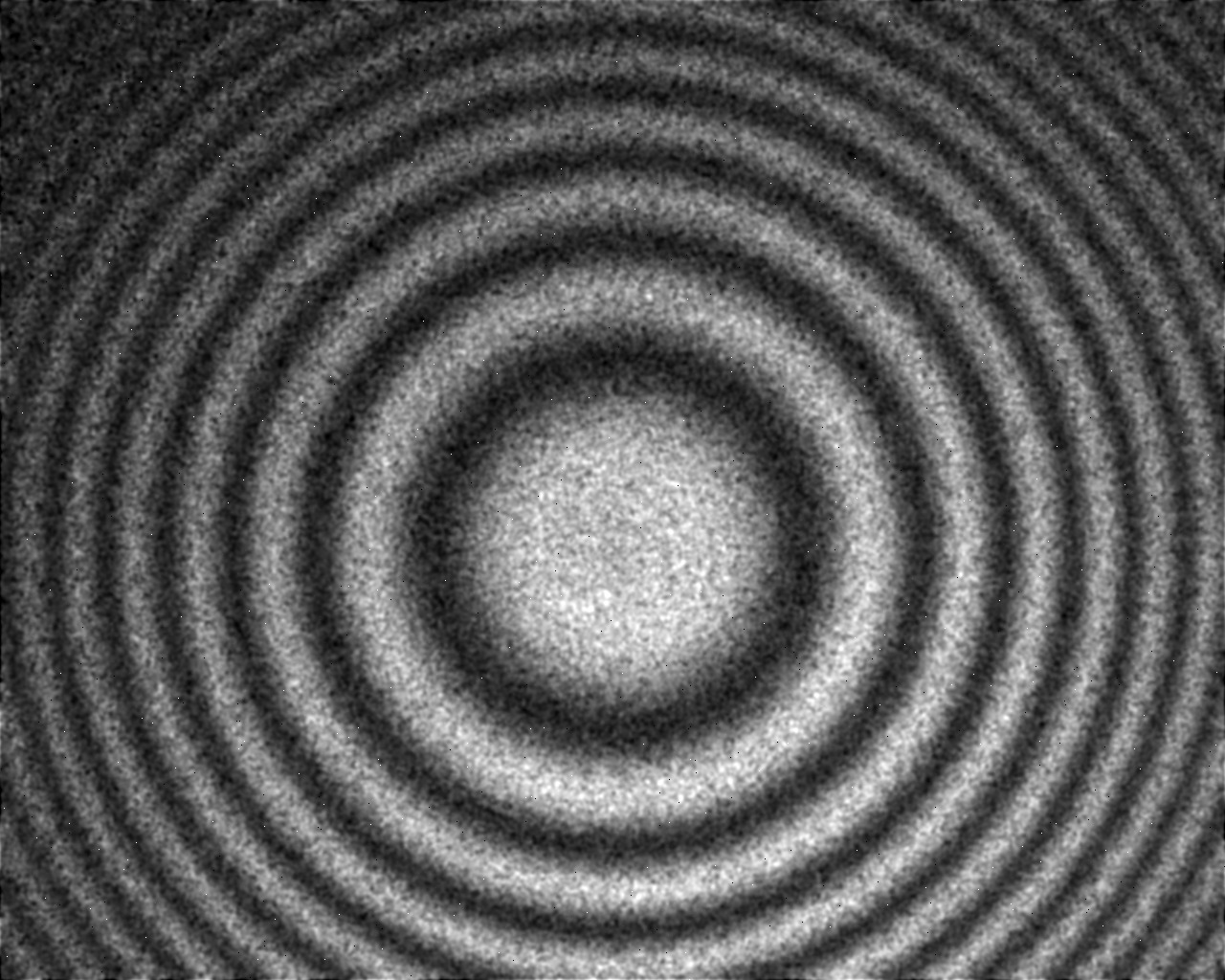}\\
        \small{\(\eta\) = 10.0}
    \end{minipage}
    
    \vspace{1em}
    
    \begin{minipage}[b]{0.45\textwidth}
        \centering
        \includegraphics[width=\textwidth]{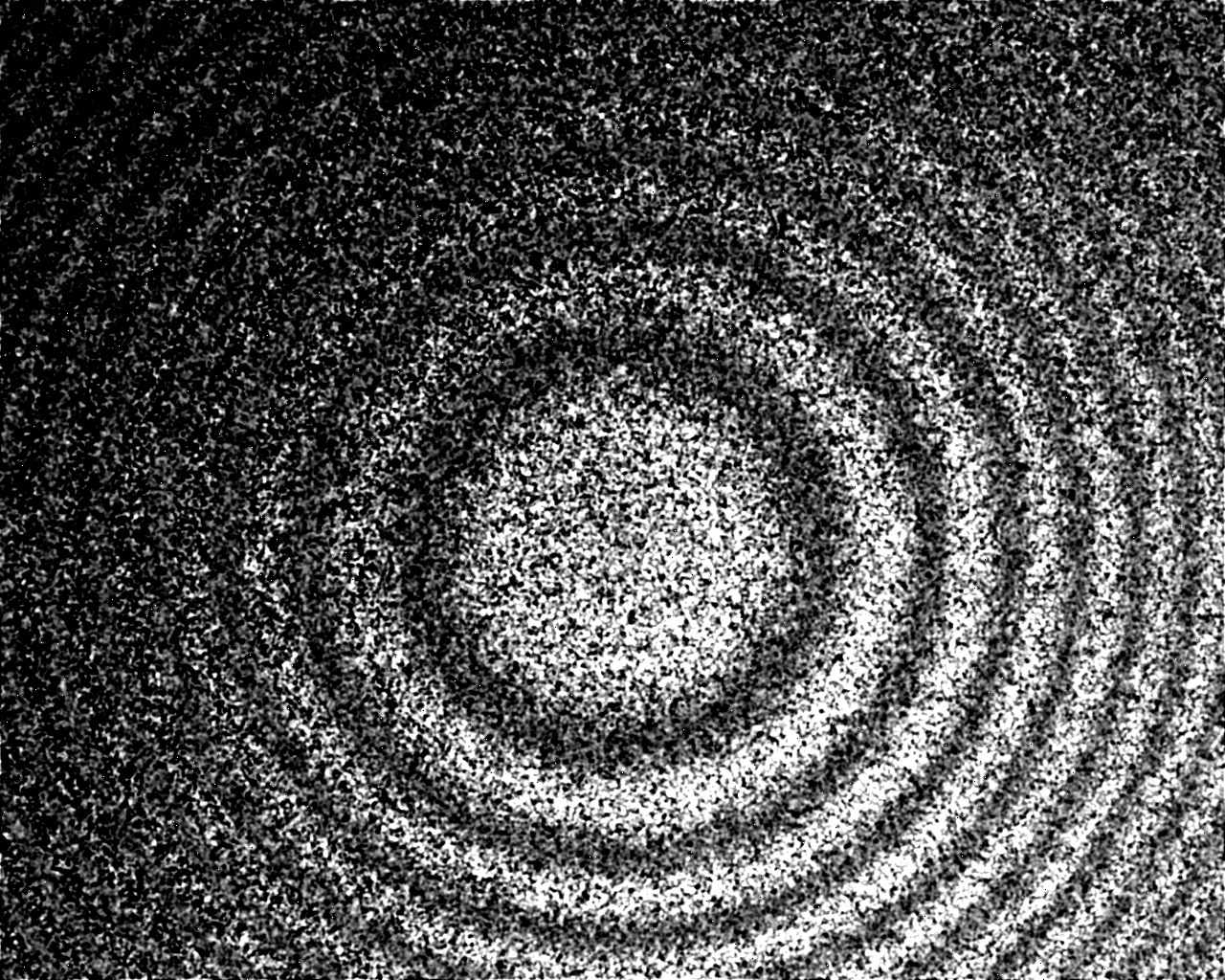}\\
        \small{\(\eta\)=30.0}
    \end{minipage}
    \hfill
    \begin{minipage}[b]{0.45\textwidth}
        \centering
        \includegraphics[width=\textwidth]{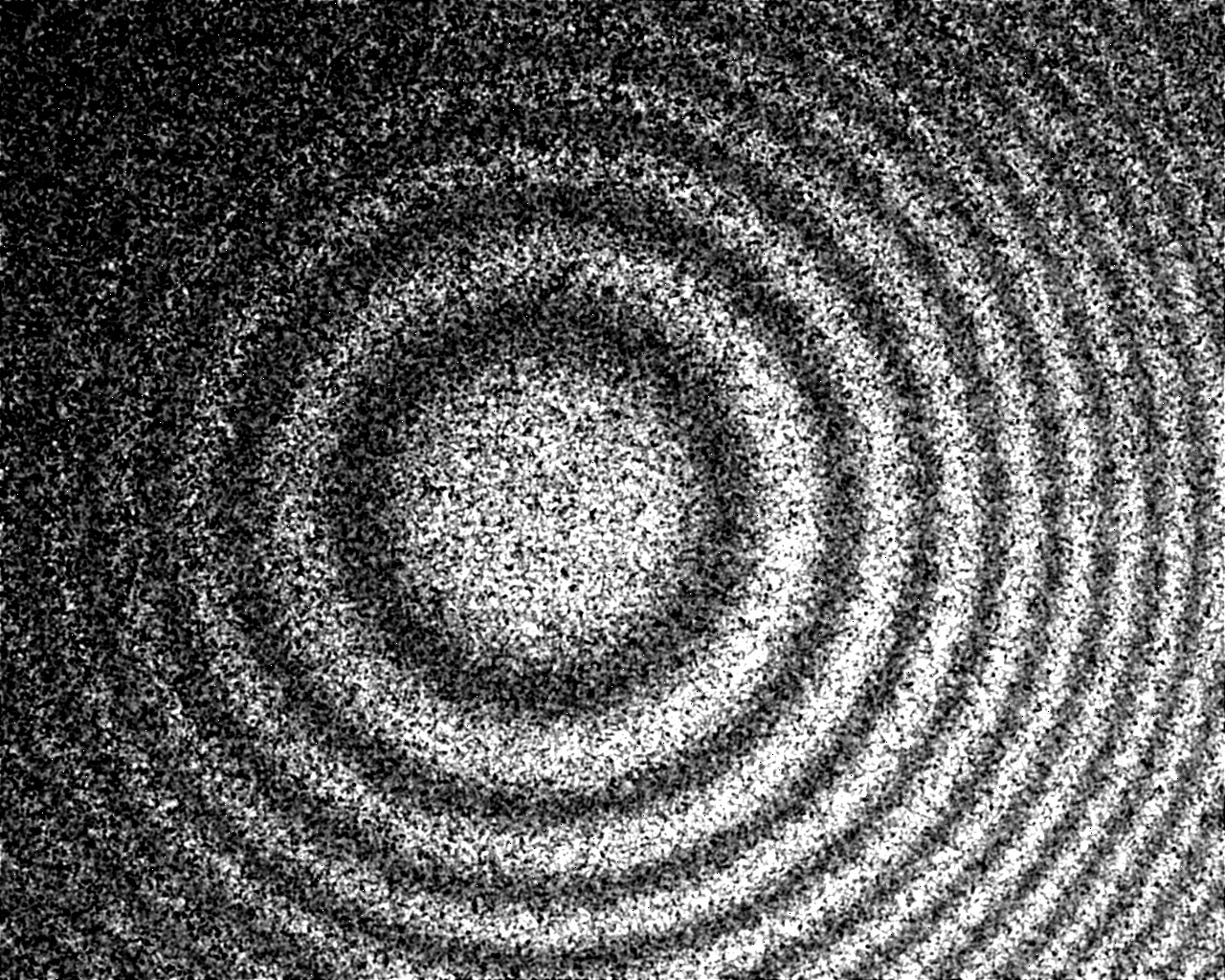}\\
        \small{\(\eta\) = 40.0}
    \end{minipage}
    
    \caption{The image demonstrates different \(eta\) values. At \(\eta\) values of 30 and 40, the noise intensity has reached the upper limit of the original image level in the experiment, and it can be expected to exceed the noise intensity in most cases}
    \label{fig:noise_change}
\end{figure}

\section{Results and Discussion}

\subsection{Performance under Different Noise Conditions}

Excluding the LSTM branch, which serves as the noise-related encoder during fine-tuning, the remaining image-based components of the network contain approximately eight million trainable parameters. 
To ensure stable convergence and physically meaningful representation learning, we generated about $2\times 10^5$ simulated interferograms as the training and validation sets for pretraining. 

In the simulation program, the range was obtained using the original fitting formula (\ref{eq:simulate}), based on three times the standard deviation of parameters derived from the experimental image . Subsequently, simulated images were generated through randomization and by incorporating noise sources. The total illumination was calculated as: 
\begin{equation}
    I'(x, y) = \beta(x,y)[[1 + D(x, y)] \, I(x, y) + n_P(x,y) + n_S(x, y)] + n_D(x, y)
\end{equation}
where \( I(x, y) \) denotes the ideal (noise-free) interference intensity pattern generated by the optical system, and \( I'(x, y) \) represents the actual measured illumination captured by the detector.

The parameter \(\eta\) controls the overall noise intensity. After \(\eta\) modulation, each simulated image is generated using the following formula:
\begin{equation}
    I_{\mathrm{final}} = I + \eta \times (I' - I)
\end{equation}
It should be noted that \(\eta\) also exhibits a clear correspondence with the signal-to-noise ratio (SNR): qualitatively, a higher \(\eta\) value corresponds to a lower SNR. The precise quantitative relationship is detailed in  supplemental Document Section 5.

After pretraining, the model achieved a validation displacement accuracy of approximately \textit{4.84$\pm$0.15}~nm (the assessment methods are described in the Supplemental Document Section 6), demonstrating that the regression branch successfully captured the sub-wavelength mapping from interferometric features to micro-displacements.

To further investigate the robustness and inference efficiency of the proposed MFN, we comprehensively modeled multiple realistic noise sources, including Gaussian noise, Poisson noise, fringe-intensity center drift, and random phase perturbations. 
A unified noise intensity parameter, denoted as $\eta$, was introduced to control the overall level of image degradation (examples of interferograms under different $\eta$ values are shown in Fig.~\ref{fig:noise_change}). 
We selected nine discrete noise intensities uniformly distributed over the range $\eta \in [0, 40]$ to quantitatively evaluate the anti-noise capability and inference speed of MFN, and to compare its performance with that of HAA.

\begin{figure}[htbp]
    \centering
    \includegraphics[width=1.0\linewidth]{ 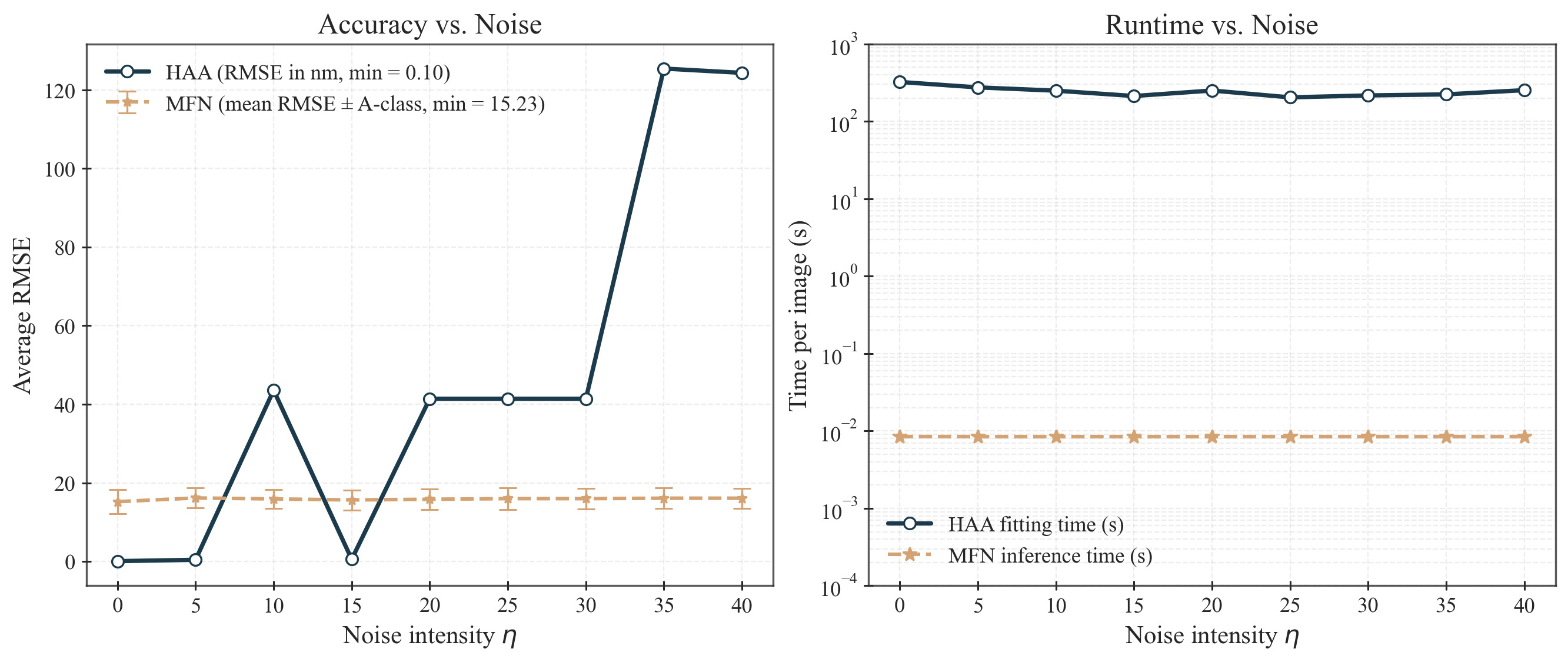}
    \caption{ \textbf{Left:} Average displacement prediction accuracy (RMSE in nanometers) versus noise intensity. 
    Due to the high computational cost of the HAA, its performance was evaluated on sampled interferogram subsets for each noise level. 
    In contrast, the MFN was evaluated on all available sequences at every $\eta$, enabling the estimation of Type-A uncertainties shown as error bars. 
    \textbf{Right:} Inference time per image versus noise intensity. 
    The MFN model was trained and tested on an NVIDIA L40 GPU. 
    For the HAA, the iterative optimization process was capped at a maximum runtime of $10^3$\,s per image to prevent unbounded fitting time; without this restriction, convergence failures frequently occurred at high noise intensities, with runtimes exceeding $10^4$\,s and no valid solution found. 
    The results highlight that the MFN maintains stable precision and achieves over four orders of magnitude acceleration compared with the traditional HAA, even under severe noise degradation.}
    \label{fig:noise_comparison}
\end{figure}

As illustrated in Fig.~\ref{fig:noise_comparison}, while the HAA achieves sub-nanometer fitting accuracy at low noise levels, its performance rapidly deteriorates as $\eta$ increases, exceeding 100~nm RMSE when $\eta=40$, primarily due to unstable phase-fitting under high noise contamination. 
In contrast, the MFN maintains stable regression accuracy across all noise intensities, with an average prediction error of about 16~nm even at the highest $\eta$. 
Regarding inference efficiency, the MFN requires only a single forward pass per image, whereas the HAA depends on iterative heuristic optimization. 
Consequently, the MFN achieves over four orders of magnitude speedup in per-frame processing, demonstrating its suitability for high-throughput or real-time interferometric applications.

\subsection{Effect of Fine-tuning Data Volume on Classification Performance}

In practical deployment, the amount of real data available for fine-tuning can vary significantly depending on experimental conditions and acquisition cost. 
In this section, we focus on the performance of the \textit{order classification head}, which is responsible for predicting the discrete interference order. 
This task provides a more direct indicator of the model’s domain adaptation capability than absolute displacement regression, since accurate order recognition reflects the network’s ability to capture essential interference semantics under real noise and contrast variations. 
Therefore, the classification accuracy serves as the primary metric for assessing fine-tuning effectiveness across different dataset sizes.

\begin{figure}[htbp]
    \centering
    \includegraphics[width=0.85\textwidth]{ 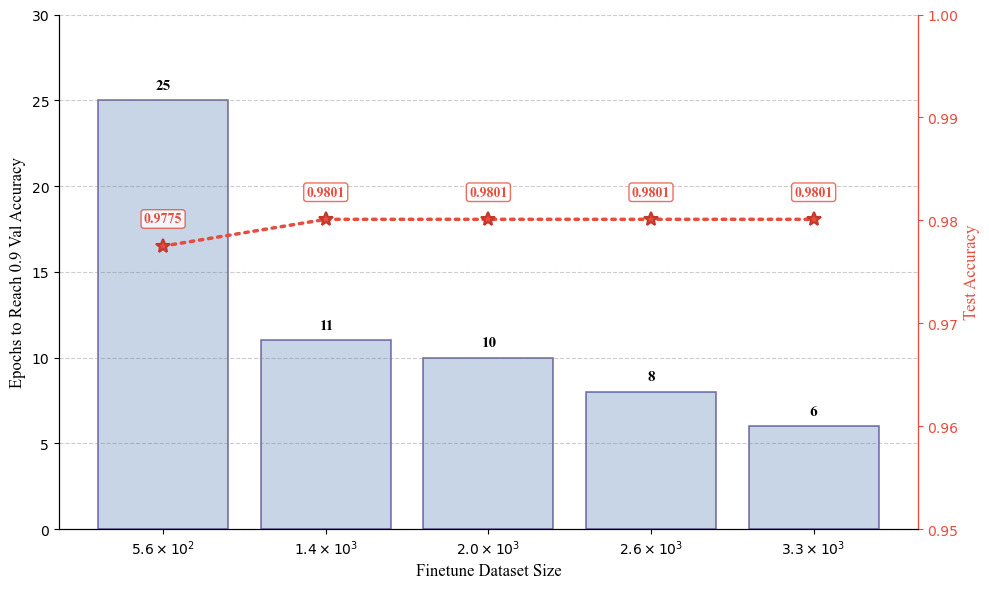}
    \caption{Influence of fine-tuning dataset size on the convergence speed and classification accuracy of the order prediction head.
    The horizontal axis represents the number of fine-tuning images, and the left vertical axis (blue bars) indicates the number of epochs required to reach 0.9 validation accuracy. 
    The red curve (right axis) shows the final test accuracy on a fixed test set of $2.5\times10^3$ images. 
    All experiments were conducted on the same hardware platform, and each fine-tuning process completed within 10–15 minutes.}
    \label{fig:data_volume}
\end{figure}

To quantitatively evaluate how data volume affects convergence behavior and final performance, we fine-tuned the pretrained model using subsets of different sizes drawn from the real interferometric dataset. 
The horizontal axis in Fig.~\ref{fig:data_volume} denotes the number of fine-tuning images, and both validation and test accuracies were measured on a fixed test set containing $2.5\times10^3$ images. 
All experiments were performed under identical hardware and hyperparameter settings, and each fine-tuning process completed within 10–15 minutes.

As shown in Fig.~\ref{fig:data_volume}, the model achieves 0.9 validation accuracy within fewer than 30 epochs once the fine-tuning dataset size exceeds approximately $5\times10^2$ images, corresponding to about 0.24\% of the pretraining dataset. 
When the data size increases further, the number of epochs required for convergence drops rapidly from 25 to fewer than 6, while the test accuracy stabilizes around 98\%. 
This demonstrates that even a small number of real interferograms are sufficient to adapt the classification head to real experimental conditions, significantly reducing the calibration cost and data acquisition burden. 
Such results also indicate that the high-level decision boundaries learned from simulated data generalize effectively to real-world interferometric patterns, validating the robustness and transferability of the proposed multimodal architecture.

From an application perspective, this data efficiency is particularly advantageous in industrial or laboratory setups where collecting large volumes of labeled interferograms is impractical. 
The ability to achieve near-saturated accuracy with only a few hundred samples highlights the practicality of the MFN framework for rapid deployment and online adaptation across diverse optical systems.

\subsection{Ablation Studies}
To verify the necessity of each core module in our three-branch multimodal architecture, and to confirm that the performance gains of MFN stem from effective complementary feature fusion rather than increased model capacity, we conducted systematic ablation studies following the training and evaluation protocols in Section 2.4. We quantified the impact of ablating key components (the MobileViT global branch, parallel CNN local branch, and LSTM temporal branch) on both displacement regression precision and interference order classification accuracy.

\begin{figure}[htbp]
    \centering
    \includegraphics[width=1\textwidth]{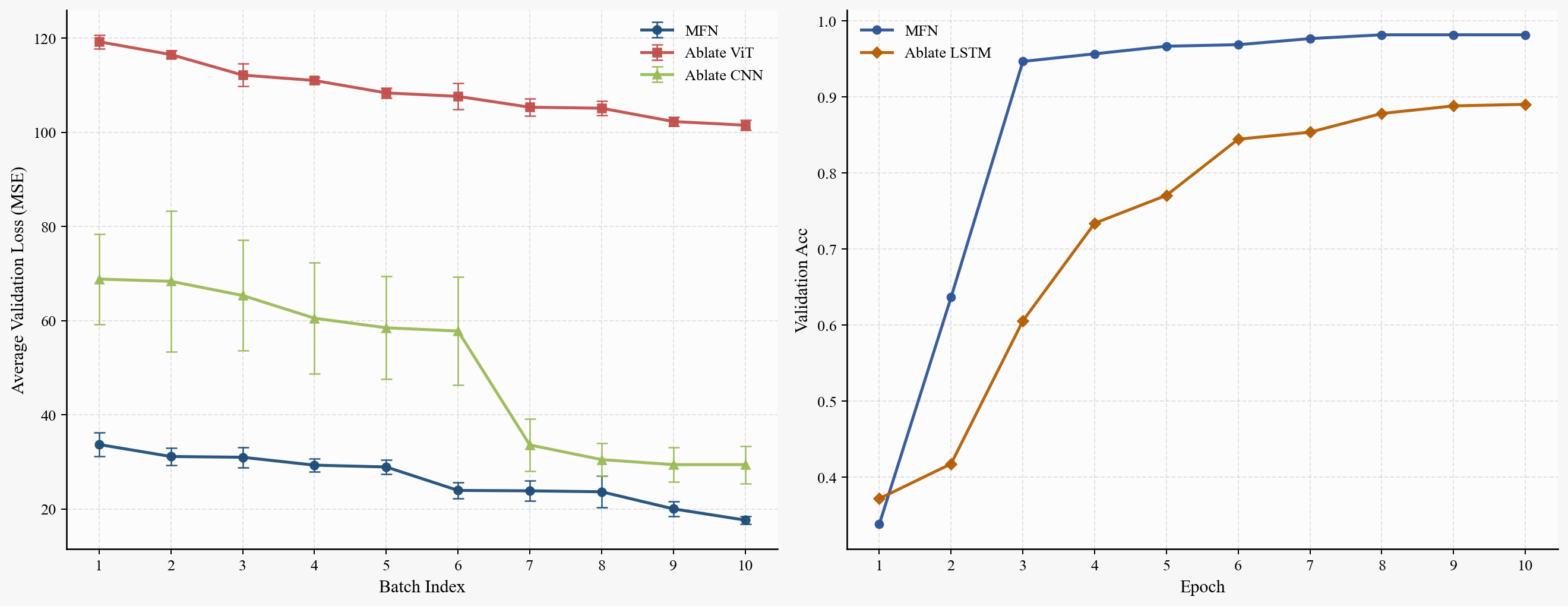}
    \caption{Ablation study results of key components in the proposed MFN architecture. (Left) Comparison of average validation MSE loss on the simulated interferogram validation set during the pretraining stage for the full MFN model, the model with MobileViT branch ablated (Ablate ViT), and the model with parallel CNN branch ablated (Ablate CNN). The x-axis denotes the Batch Index: to optimize video memory usage during pretraining, the full simulated dataset is divided into 10 batches with consistent data distribution (interferograms with different noise intensities are uniformly distributed across all batches), and each batch is trained for 20 epochs. Error bars represent one standard deviation of the measurement results. (Right) Comparison of validation order classification accuracy during the fine-tuning stage for the full MFN model and the model with LSTM temporal branch ablated (Ablate LSTM). The LSTM branch is designed to capture the statistical relationship between temporal noise evolution and interference order transitions, thus classification accuracy is adopted as the primary evaluation metric for this ablation experiment.}
    \label{fig:ablation}
\end{figure}

As shown in Fig. \ref{fig:ablation} (left), the full MFN model achieves the lowest validation MSE across all batches, converging to a final loss of 17.8 $\text{nm}^2$ (corresponding to a regression precision of $\approx$ 4.22 nm) with minimal training fluctuation. Ablating the MobileViT branch leads to the most severe performance degradation, with the final MSE rising to 101.5 $\text{nm}^2$. The model with the CNN branch ablated exhibits drastically larger training fluctuations and wider error bars, despite a lower initial loss than the CNN-only variant, indicating poor convergence stability.

This performance gap arises from the intrinsic complementarity of the two feature extraction backbones. CNNs have an inherent local spatial inductive bias, enabling efficient extraction of the fine-grained fringe phase structures critical for sub-wavelength displacement regression. The MobileViT backbone, by contrast, models long-range global spatial dependencies via self-attention, providing global contextual constraints to suppress noise-induced local feature distortion \cite{mehta2022mobilevitlightweightgeneralpurposemobilefriendly, liu2026lightweight}. Removing the MobileViT branch deprives the model of global fringe distribution information, causing severe feature drift under high noise; removing the CNN branch eliminates the local texture inductive bias, leading to unstable feature learning on noisy data, consistent with reported limitations of transformer-only architectures in precision optical measurement tasks \cite{liu2026lightweight, maurya2025attention}. The full MFN fuses these complementary features, achieving both fine-grained phase extraction and global context constraint for optimal regression performance and stability.

For the fine-tuning stage, we ablated the LSTM temporal branch, which is activated exclusively during fine-tuning to capture temporal statistical cues for order classification. As shown in Fig. \ref{fig:ablation} (right), the full MFN reaches over 95\% classification accuracy at the 3rd epoch and converges to 98.1\% final accuracy, while the LSTM-ablated model converges significantly slower and only reaches 88.9\% final accuracy (a ~10 percentage point drop). This confirms that the LSTM branch provides critical temporal dependency information of fringe statistical features across consecutive frames, which cannot be captured by single-frame image branches. Without this temporal context, the model suffers frequent order misjudgments under real experimental noise.

In summary, our ablation studies confirm that the MFN’s superior performance comes not from increased parameter count, but from effective multi-modal feature fusion via the multi-branch architecture. The CNN, MobileViT, and LSTM branches each deliver irreplaceable feature representations, whose synergy ensures the model’s excellent regression and classification performance, fully validating the necessity of our multimodal design.

\section{Conclusion}

In this work, we proposed a multimodal fusion network (MFN) to address the intrinsic $\lambda/2$ phase ambiguity that constrains single-wavelength interferometric displacement measurement. 
By integrating image-based features and temporal statistical cues within a dual-head architecture, the model performs sub-$\lambda/2$ displacement regression and integer-order classification simultaneously, yielding a unified and physically interpretable displacement estimation framework. 
Unlike traditional dual-wavelength approaches that require high signal quality and complex optical setups, MFN resolves the $\lambda/2$ ambiguity purely through learning-based inference, maintaining accuracy and stability even under strong noise or fringe degradation. 
This data-driven strategy enables precise displacement prediction using a single optical channel, reducing both hardware complexity and cost.

Experimental evaluations demonstrate that the MFN achieves a regression precision of $4.84\pm0.15$~nm using $2\times10^{5}$ simulated interferograms for pretraining, and reaches a classification accuracy of 98\% on real test data after fine-tuning with only 0.24\% of the pretraining dataset (approximately $5\times10^{2}$ real interferograms). 
Under severe mixed-noise conditions, the proposed model maintains stable displacement accuracy of about 16~nm, whereas the conventional Heuristic Analytical Algorithm (HAA) exhibits fitting errors exceeding 100~nm. 
Moreover, the MFN completes inference in approximately 10 ~ms per image, over four orders of magnitude faster than HAA, which typically requires hundreds of seconds per frame. 
These results confirm that the proposed framework effectively combines precision, robustness, and computational efficiency, enabling scalable, real-time interferometric metrology for both research and industrial applications.

Overall, the MFN offers a data-driven, hardware-efficient, and noise-tolerant solution for nanoscale displacement sensing. 

Meanwhile, current performance is mainly limited by the calibration accuracy of the voltage–displacement (V–d) measurement device, which mainly influences the accuracy of displacement head when fine tuning(Detailed error and accuracy analysis can be found in the Supplemental Document). 
Due to the range limitations of the calibration method, the  study mainly focuses on the displacement range of 0–2 $\mu m$. Within this range, the MFN achieves an accuracy of $98\%$ and exhibits strong robustness. In principle, the proposed framework can be extended to measure larger displacements; however, such measurements may be affected by more types of noise, which would require different parameter settings and optimization strategies.
Future work will  be improved in system calibration, particularly in voltage–displacement (V–d) precision, and focus on extending this architecture to other forms of phase-related measurement, such as angular, thermal, and material property sensing. 


\section*{CRediT authorship contribution statement}
\textbf{Zixing Jia:} Conceptualization, Data curation, Formal analysis, Investigation, Methodology, Project administration, Software, Validation, Visualization, Writing - original draft, Writing - review \& editing. \textbf{Jiawei Li:} Conceptualization, Data curation, Formal analysis, Investigation, Methodology, Software, Validation, Visualization, Writing - original draft, Writing - review \& editing. \textbf{Xin Li:} Funding acquisition, Resources, Supervision, Writing - review \& editing. \textbf{Kuang Zheng:} Investigation, Writing - review \& editing. \textbf{Yuehua Chen:} Investigation. \textbf{Ziping Chen:} Funding acquisition, Resources. \textbf{Xudong Lin:} Funding acquisition, Resources, Writing - review \& editing.

\section*{Supplemental Document}
 See Supplemental Document for supporting content.

\section*{Funding}
National Key Research and Development Program of China (2022YFC2203800)

\section*{Declaration of competing interest} 
The authors declare that they have no known competing financial 
interests or personal relationships that could have appeared to influence 
the work reported in this paper.

\section*{Acknowledgment}

We acknowledge Guangdong Provincial Demonstration Center project of
Experimental Education in 2023 (The Center of Experimental Physics and Astronomy Education) for assistance.

\section*{Data availability}
Data will be made available on request.


\bibliographystyle{elsarticle-num} 
\bibliography{references}






\clearpage

\section*{Table:}
\begin{table}[ht]
\centering
\small
\renewcommand{\arraystretch}{2.3}
\begin{tabularx}{\textwidth}{>{\raggedright\arraybackslash}p{2.6cm} | >{\raggedright\arraybackslash}X | >{\raggedright\arraybackslash}p{5.8cm}}
\hline
\textbf{Category} & \textbf{Feature Name} & \textbf{Formula} \\

\hline
\multirow{3}{*}{\textbf{Image Channels}} 
& Original Interferogram & $I_{\text{orig}}(x,y) = I_t(x,y)$ \\

\cline{2-3}
& Frame Difference Image & $I_{\text{diff}}(x,y) = |I_t(x,y) - I_{t-1}(x,y)|$ \\

\cline{2-3}
& Frequency-Domain Image &
$ I_{\text{fft}}(u,v) =
\dfrac{\log\!\left(|\mathcal{F}[I_t](u,v)| + \epsilon \right) -\min_{u,v}}{\max_{u,v} - \min_{u,v}}
$ \\

\hline
\multirow{7}{*}{\textbf{Numerical Channels}} 
& Normalized Frame Index & $f_1 = \dfrac{t}{T-1}$ \\

\cline{2-3}
& Mean Intensity & $f_2 = \dfrac{1}{HW} \sum_{x,y} I_t(x,y)$ \\

\cline{2-3}
& Intensity Variance & $f_3 = \dfrac{1}{HW} \sum_{x,y} ( I_t(x,y) - f_2 )^2$ \\

\cline{2-3}
& Number of Horizontal Fringe Peaks &
$f_4 = \texttt{PeakCount}(P(y)),\quad P(y) = \dfrac{1}{H}\sum_x I_t(x,y)$ \\

\cline{2-3}
& Low-Frequency Energy Ratio &
$f_5 = 
\dfrac{
\sum_{(u,v)\in M} |\hat{I}(u,v)|^2
}{
\sum_{u,v} |\hat{I}(u,v)|^2
}$ \\

\cline{2-3}
& Spectral Centroid Distance &
$f_6 =
\sqrt{
(\bar{u}-u_0)^2 + (\bar{v}-v_0)^2
},
\quad
\bar{u} = 
\dfrac{
\sum_{u,v} u\,\hat{I}(u,v)
}{
\sum_{u,v} \hat{I}(u,v)
}
$ \\

\cline{2-3}
& Image Entropy &
$f_7 = -\sum_{i=1}^{256} p_i \log_2 (p_i + \delta)$ \\

\hline
\end{tabularx}
\caption{Feature representation and temporal--spatial encoding. Key symbols: $I_t(x,y)$ is the interferometric frame at time $t$; $H,W$ are image dimensions; $T$ is the sequence length; $\mathcal{F}$ denotes the 2D FFT; $\epsilon,\delta$ are stability constants; $M$ is the low-frequency mask; $(u_0,v_0)$ denotes the spectral center; $p_i$ is the normalized histogram. Further details and justification of each feature are provided in the Supplemental Document.}
\label{tab:features}
\end{table}

\begin{table}[ht]
\centering

\begin{tabularx}{\textwidth}{>{\raggedright\arraybackslash}p{2.2cm} | >{\raggedright\arraybackslash}X | >{\raggedright\arraybackslash}p{3.2cm}}
\hline
\textbf{Symbol} & \textbf{Meaning} & \textbf{Parameter Settings} \\

\hline
\( D(x, y) \) & \textbf{Power Drift }:
                \( D(x, y) = 1-\xi, \quad \xi \sim \mathcal{N}(1, \sigma_d^{2}) \) & \(\sigma_d = 0.005\) \\

\hline
\( n_{P1}(x, y) \) & \textbf{Photon Shot Noise}:
                \( n_{P1}(x, y) \sim \mathrm{Poisson}([1 + D(x, y)] \, I(x, y) )\) &  \(n_1 =n_{P1}  + 2.5\times \mathcal{U}(1.2,1.8)\) \\

\hline
\( n_P(x, y) \) & \textbf{PSF}:
                \( n_P(x, y) \sim \mathcal{N}(n_1, \sigma_R^2) \) &\(\sigma_R = \mathcal{U}(1.2,2.0)\)\\

\hline
\( n_S(x, y) \) & \textbf{Stripe-Type Structured Noise}:
                \( n_S(x, y) = A_s \sin\left( 2\pi \dfrac{y}{\lambda_s} + \delta(x,y) \right) \) & \(A_s\): \(\mathcal{U}[0.6S, 0.9S]\), \(S=1.8\)\(\lambda_s\): \(\mathcal{U}[8, 20]\) pixels
                \(\delta\): \(\mathcal{N}(0, 0.2^2)\)\\

\hline
\( \beta(x, y)\) & \textbf{Background Gradient + Vignetting}:\(\beta =(1 - v(x^2 + y^2))(1+\alpha x + \delta y + \gamma(xy))\) &  \(\alpha = 0.1\quad \delta = 0.08\quad  \gamma = 0.05 \quad v = 0.005\)\\

\hline
\( n_{D}(x, y) \) & \textbf{Discrete Noise}:salt-and-pepper noise and dead pixel &  dead pixel ratio =0.002 salt-and-pepper ratio =  0.004\\

\hline
\end{tabularx}
\caption{\textbf{Noise Models and Parameter Settings}}
\label{tab:noise_models}
\end{table}

\end{document}